\documentclass[10pt,letterpaper]{article}
\usepackage[top=1in,bottom=1in,left=1.5in,right=1in]{geometry}

\usepackage{changepage}
\usepackage{amsmath,amssymb}
\usepackage{cite}
\usepackage{nameref,hyperref}
\usepackage{setspace} 
\usepackage{graphicx}

\makeatletter
\renewcommand{\@biblabel}[1]{\quad#1.}
\makeatother

\usepackage{color}

\begin{document}

\begin{center}
{\huge Linking in domain-swapped protein dimers}
\end{center}

Marco Baiesi$^{1,2,\*}$,
Enzo Orlandini$^{1,2}$,
Antonio Trovato$^{1,3}$,
Flavio Seno$^{1,3}$,
\\
{\bf 1} Department of Physics and Astronomy, University of Padova, Padova, Italy
\\
{\bf 2} INFN - Sezione di Padova, Padova, Italy
\\
{\bf 3} CNISM, Padova, Italy
\\
* E-mail: Corresponding baiesi@pd.infn.it

\section*{Abstract}
The presence of knots has been observed in a small fraction of single-domain proteins and related to their thermodynamic and kinetic properties. The exchanging of identical structural elements, typical of domain-swapped proteins, make such dimers suitable candidates to validate the possibility that mutual entanglement between chains may play a similar role for protein complexes. We suggest that such entanglement is captured by the linking number. This represents, for two closed curves, the number of times that each curve winds around the other. We show that closing the curves is not necessary, as a novel parameter $G'$, termed Gaussian entanglement, is strongly correlated with the linking number. Based on $110$ non redundant domain-swapped dimers, our analysis evidences a high fraction of chains with a significant intertwining, that is with $|G'| > 1$. We report that Nature promotes configurations with negative mutual entanglement and surprisingly, it seems to suppress intertwining in long protein dimers. Supported by numerical simulations of dimer dissociation, our results provide a novel topology-based classification of protein-swapped dimers together with some preliminary evidence of its impact on their physical and biological properties.

\section*{Introduction}
In biological systems, proteins rarely act as isolated monomers and
association to dimers or higher oligomers is a commonly observed
phenomenon~\cite{nussinov2015,mackinnon2015,marianayagam2004,deremble2005,ellis2001,ouzounis2003,worth2009,wodak2002}.
Recent structural and biophysical studies show that protein
dimerization or oligomerization is a key factor in the regulation of
proteins such as enzymes~\cite{bonafe99}, ion channels\cite{hebert98},
receptors and transcription factors~\cite{pingoud2001,beckett2001}. In
addition, this mechanism can help to minimize genome size, while
preserving the advantages of modular complex
formation\cite{marianayagam2004}. Oligomerization, however, can also have deleterious
consequences when non-native oligomers, associated with pathogenic
states, are generated\cite{viola2015,knowles2014,fandrich2012,kayed2003,shameer2012}. 
Specific protein dimerization is integral to
biological function, structure and control, and must be under
substantial selection pressure to be maintained with such frequency
in living organisms.

Protein-protein interactions may occur between identical or
non-identical chains (homo or hetero-oligomers) and the association
can be permanent or transient\cite{ali2015}. Protein complexes can widely differ
based on their affinity.  Binding affinities, evaluated for dimers
as dissociation constants, can cover up to nine orders of magnitude,
highlighting the fact that a strong modulation is necessary to hold up
the full protein interaction network\cite{kastritis2011,kastritis2013}.

The mechanisms for the evolution of oligomeric interfaces and those
for the assembly of oligomers during protein synthesis or refolding
remain unclear. Different paradigms have been proposed for the
evolution of protein oligomers, among which figures three-dimensional
(3D) domain swapping~\cite{jaskolski2013,bennet1994,bennet1995,green1995,liu2002}.

\begin{figure}[!t]
\includegraphics[width=6in]{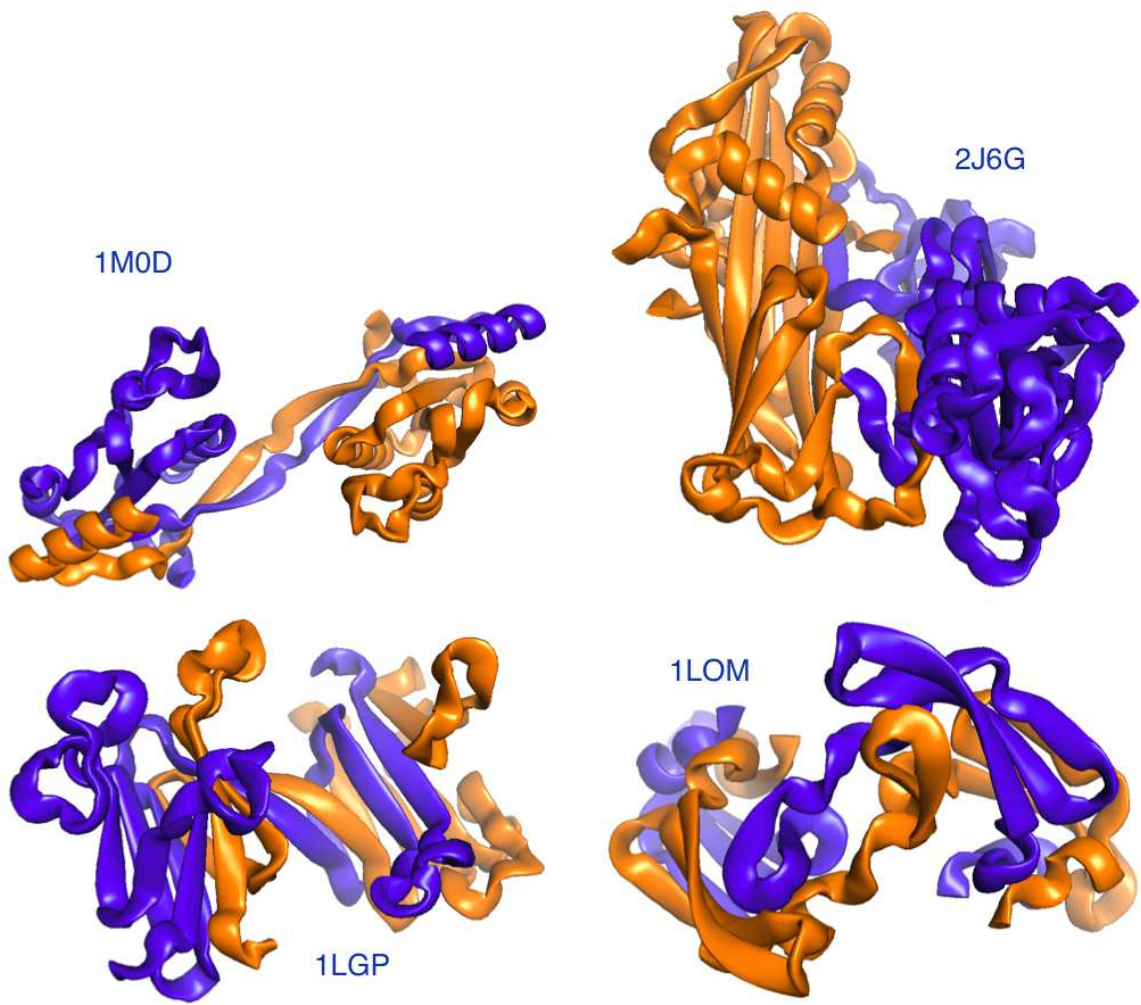} 
\caption{
{\bf Some domain-swapped dimers with high, negative linking number}}
\label{fig:ex1}
\end{figure}

\begin{figure}[!t]

\includegraphics[width=6in]{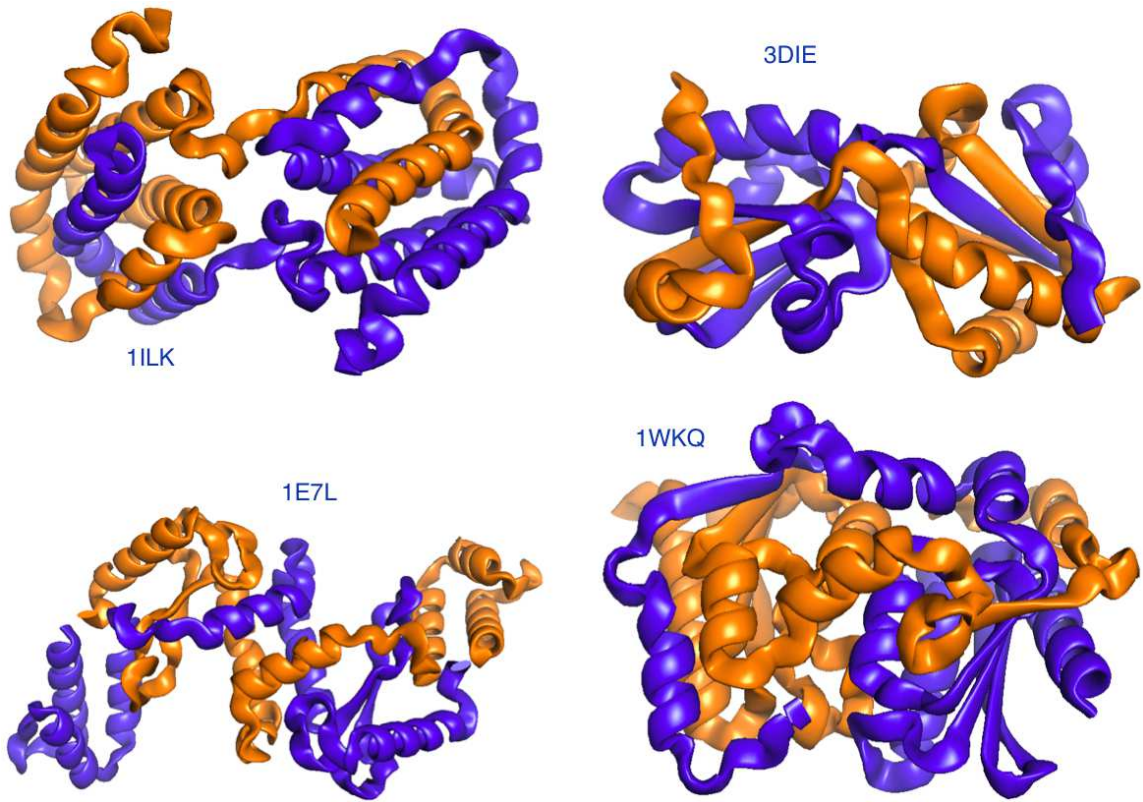} 
\caption{{\bf Some domain-swapped dimers with high, positive linking number}}
\label{fig:ex2}
\end{figure}

Three-dimensional domain swapping is a mechanism through which two or
more protein molecules form a dimer or higher oligomer by exchanging
an identical structural element (see Figures~\ref{fig:ex1} and~\ref{fig:ex2}). 
Several native (natural/physiological) intra-molecular interactions within the monomeric
structures are replaced by inter-molecular interactions of protein
structures in swapped oligomeric
conformations\cite{yang2005}. Critical in this process is the hinge
region, the only polypeptide segment that needs to adopt different
conformations in the monomer and in the domain-swapped oligomer. Domain
swapping is typically the response of the protein to relieve
conformational stress that is present in this hinge region of the
monomer.  Structures in swapped conformations were reported to perform
a variety of functions, and proteins involved in deposition diseases
(like neurodegenerative diseases, amyloidosis and Alzheimer’s disease)
have been reported in 3D domain-swapped
conformations\cite{alva2007,liu2001,jaskolski2001,rousseau2012}.

Domain-swapped proteins may assume rather
complicated spatial structures, since the swapping arms in their
rotation can wind up forming tightly compenetrated structures.
Examples are shown in Figure~\ref{fig:ex1} and Figure~\ref{fig:ex2}.
For instance, the Staphylococcus aureus thioredoxin (Trx) fold mutant (pdb
code 3DIE) represented in Figure~\ref{fig:ex2}
clearly illustrates the deep clinging between the two proteins.

The apparent intertwining  between proteins assembled in complexes 
is certainly a distinguishing characteristic of these systems.
An interesting issue to explore is the possibility of 
introducing topology-based descriptors that can capture the
entanglement in a robust and measurable way, and that can be 
related to either some physico/chemical properties or biological functionalities.

For instance, for single
chain globular proteins it has been observed that the backbones may
entangle themselves into a physical knot 
~\cite{mansfield1997,taylor2000,virnau2006,lua2006,jamroz2014,lim2015} that does not disentangle 
even in the denatured unfolded state~\cite{mallam2010}. Knotted proteins are
interesting because they are rare, and their folding mechanisms
and function are not well understood, although it has been proposed
that they might increase thermal and kinetic stability~\cite{sulkowska2008,mallam2008,soler2013}.

Although knots are mathematically defined 
only for closed loops~\cite{rolfsen1976}, in the last decade there have been 
several attempts to introduce sufficiently robust and topologically inspired measures of 
knots in open chains~\cite{orlandini2007}. For a single protein, for instance, 
one can close its backbone by connecting  the N and C termini to a point (chosen randomly)
on the surface of a sphere that contains the chain. This sphere can be either  very 
large compared to the chain size (closure at infinity)
or it can be replaced by the convex hull of the chain.
In all cases the artificial closure can introduce additional entanglement and 
there have been suggested several ways to either mitigate or control 
this problem~\cite{janse1992,marcone2005,millet2005,tubiana2011}

For two proteins forming a dimer, if one is interested in measuring the degree 
of mutual entanglement, the notion of a knotted open chain must be generalised to 
that of a link between  two open chains. In analogy with  the procedure 
used for knots, one can think of closing artificially the two backbones
to generate two loops.
This can be done by 
joining the ends of each protein at infinity and computing a link 
invariant, such as the linking number $\cal G$, an integer index that

quantifies the number of times that  a curve winds around the other
(technically, with Gauss integrals evaluated over the
protein backbone, $\cal G$ detects the degree of homological linking 
between two closed curves~\cite{rolfsen1976} and can be used
to classify proteins~\cite{fain2004}).

Two curves are not linked if ${\cal G}=0$  while ${\cal G}=1$ 
or ${\cal G}=-1$ denotes the simplest link between two loops.
Since the sign of ${\cal G}$ depends on the orientation of the two curves,
in our implementation we choose to follow the standard N-C orientation 
along the backbones of the proteins.
As in knots, the random closure may introduce additional linking
between the two chains and an estimate of ${\cal G}$ is necessarily a probabilistic one 
that requires a sufficiently large number of closure paths~\cite{millet2005}. 

We denote by $G$ the average of the linking number $\cal G$ over many closures.
Along with  $G$, we consider a 
{\it Gaussian entanglement} indicator, $G'$, 
computed with the same method adopted for $G$ but
without closing the curves.

We show that $G'$ strongly correlates with its topological
counterpart $G$.  
Both estimators are used to analyse a non-redundant
set of 3D domain swapped dimers. It turns out that several dimers
present a high degree of mutual entanglement.

The short CPU time required to estimate $G'$ 
allows a quick systematic mining of linked dimers from protein databanks.
With a computationally much heavier test, for some dimers
we check whether this measure of entanglement is robust or is just an
artifact of the specific crystallographic structure found in the
database. This is done by performing simplified molecular
dynamics simulations, starting from several native structures with
non-trivial values of $G'$. The time evolution of $G'(t)$
during the process of dissociation of the dimer reveals additional
information on the amount of intertwining of the two proteins.

\section*{Results}

\begin{table}
\begin{tabular}{| lllrl | lllrl | lllrl |}
\hline
{\bf dimer} & $G'$ & $G$ & $N$ && {\bf dimer} & $G'$ & $G$ & $N$  && {\bf dimer} & $G'$ & $G$ & $N$  &\\
\hline 
1M0D & -2 & -1.65 & 129 &  & 1K51 & -0.35 & -0.47 & 72 &  & 1BJ3 & 0.02 & 0 & 129 &  \\
2J6G & -1.79 & -2.08 & 260 &  & 2A62 & -0.35 & -0.13 & 319 &  & 2QYP & 0.02 & 0 & 78 & h \\
2XDP & -1.75 & -2.23 & 123 & h & 1N9J & -0.35 & -0.32 & 98 & h & 1X2W & 0.03 & 0.01 & 129 &  \\
2Z0W & -1.67 & -1.79 & 72 & h & 1K4Z & -0.33 & -0.42 & 157 &  & 1QB3 & 0.05 & 0 & 113 &  \\
1I1D & -1.42 & -1.26 & 156 &  & 1TIJ & -0.33 & -0.22 & 114 & h & 1AOJ & 0.09 & 0.29 & 60 &  \\
2P1J & -1.34 & -1.42 & 164 &  & 1ZVN & -0.32 & -0.13 & 99 &  & 1S8O & 0.1 & 0.23 & 545 & h \\
1LGP & -1.32 & -1.28 & 113 & h & 2A4E & -0.28 & -0.25 & 208 &  & 1CDC & 0.13 & 0.23 & 96 &  \\
1NPB & -1.3 & -1.08 & 140 &  & 1DXX & -0.27 & -0.17 & 238 & h & 1WY9 & 0.16 & -0.03 & 111 &  \\
1LOM & -1.22 & -1.45 & 101 &  & 1K04 & -0.27 & -0.63 & 142 & h & 3NG2 & 0.17 & 0.25 & 66 &  \\
2BZY & -1.19 & -1.43 & 62 & h & 3FJ5 & -0.2 & -0.23 & 58 &  & 2HZL & 0.19 & 0.59 & 338 &  \\
1KLL & -1.16 & -0.78 & 125 &  & 1CTS & -0.17 & -0.5 & 437 &  & 2FPN & 0.2 & 2.72 & 198 &  \\
1HW7 & -1.13 & -1.44 & 229 &  & 1FOL & -0.17 & -0.35 & 416 &  & 2CN4 & 0.26 & 0.6 & 173 &  \\
1BYL & -1.06 & -1.12 & 122 &  & 2BI4 & -0.15 & -0.02 & 382 &  & 2SPC & 0.29 & 0.6 & 107 &  \\
1MI7 & -1.05 & -0.97 & 103 &  & 2ONT & -0.15 & -0.09 & 73 & h & 1R5C & 0.31 & 0.5 & 124 &  \\
1BUO & -1.03 & -0.99 & 121 & h & 2CI8 & -0.14 & -0.61 & 106 & h & 2CO3 & 0.34 & 0.53 & 142 &  \\
1O4W & -0.95 & -1.34 & 123 &  & 1DWW & -0.12 & -0.29 & 420 &  & 2OQR & 0.34 & 0.3 & 227 &  \\
1MU4 & -0.94 & -1.09 & 86 &  & 1QQ2 & -0.12 & -0.4 & 173 &  & 1H8X & 0.35 & 0.44 & 125 & h \\
1W5F & -0.92 & -0.48 & 315 &  & 1XMM & -0.1 & -0.32 & 288 & h & 3HXS & 0.36 & -0.14 & 120 &  \\
1FRO & -0.88 & -1.08 & 176 & h & 1Q8M & -0.09 & 0.31 & 121 & h & 1GP9 & 0.41 & 0.54 & 170 & h \\
2VAJ & -0.88 & -0.52 & 93 & h & 1GT1 & -0.08 & -0.1 & 158 &  & 2FQM & 0.49 & 0.49 & 65 &  \\
1HT9 & -0.82 & -0.87 & 76 &  & 1NNQ & -0.07 & 0.26 & 170 &  & 2DSC & 0.49 & 0.51 & 195 & h \\
3FSV & -0.79 & -0.38 & 119 &  & 3D94 & -0.07 & -0.04 & 289 & h & 1SK4 & 0.56 & 0.77 & 162 & h \\
1ZK9 & -0.78 & -0.69 & 110 &  & 2IV0 & -0.04 & -0.04 & 412 &  & 2DI3 & 0.62 & 0.49 & 231 &  \\
3LOW & -0.73 & -0.63 & 100 & h & 2W1T & -0.04 & -0.07 & 175 &  & 2NZ7 & 0.63 & 0.55 & 93 & h \\
1T92 & -0.66 & -0.61 & 108 &  & 1L5X & -0.03 & -0.04 & 270 &  & 2HN1 & 0.66 & 1 & 246 &  \\
2ES0 & -0.65 & -0.54 & 131 & h & 2E6U & -0.03 & -0.02 & 142 &  & 1OSY & 0.71 & 1 & 114 &  \\
2RCZ & -0.53 & -0.55 & 79 & h & 2O7M & -0.03 & -0.04 & 153 &  & 2A9U & 0.76 & 0.86 & 133 & h \\
1KAE & -0.51 & -0.46 & 427 &  & 4AEO & -0.03 & 0 & 353 &  & 1A2W & 0.78 & 0.83 & 124 &  \\
1NIR & -0.51 & -0.5 & 538 &  & 3PSN & -0.02 & -0.03 & 183 &  & 1QX7 & 0.81 & 1.08 & 136 &  \\
1PUC & -0.49 & -0.47 & 101 &  & 1HE7 & -0.02 & -0.06 & 114 & h & 1QX5 & 0.83 & 1.09 & 145 &  \\
1HUL & -0.49 & -0.66 & 108 & h & 1U4N & -0.01 & 0.3 & 308 &  & 1MV8 & 0.86 & 1 & 436 &  \\
1I4M & -0.48 & -0.41 & 108 & h & 1UKM & -0.01 & -0.01 & 131 &  & 1QWI & 0.93 & 1.12 & 140 &  \\
2NU5 & -0.43 & -0.63 & 123 &  & 1YGT & -0.01 & -0.12 & 104 &  & 1WKQ & 1.13 & 1.06 & 158 &  \\
3LYQ & -0.42 & -0.74 & 184 &  & 1XUZ & 0 & 0.04 & 348 &  & 1E7L & 1.26 & 1.37 & 166 &  \\
2GUD & -0.41 & -0.63 & 122 &  & 2PA7 & 0 & -0.04 & 135 &  & 3DIE & 1.5 & 1.41 & 106 &  \\
1R7H & -0.39 & -0.57 & 74 &  & 1VJ5 & 0 & 0 & 547 & h & 1ILK & 1.68 & 1.63 & 151 & h \\
1ZXK & -0.38 & -0.14 & 96 &  & 2JFL & 0 & 0 & 286 & h &  &  &  &  &  \\
\hline
\end{tabular}
\caption{{\bf Domain-swapped dimers ranked from lowest to highest Gaussian entanglement $G'$.}
The mean linking number ($G$) and the 
number of amino acids in each protein of the dimer ($N$) are also reported for the
analysed dimers. Human proteins are tagged with a ``h''.
\label{tab:G1G}
}
\end{table}

\subsection*{Databank}

\begin{figure}[!t] 
\includegraphics[width=5in]{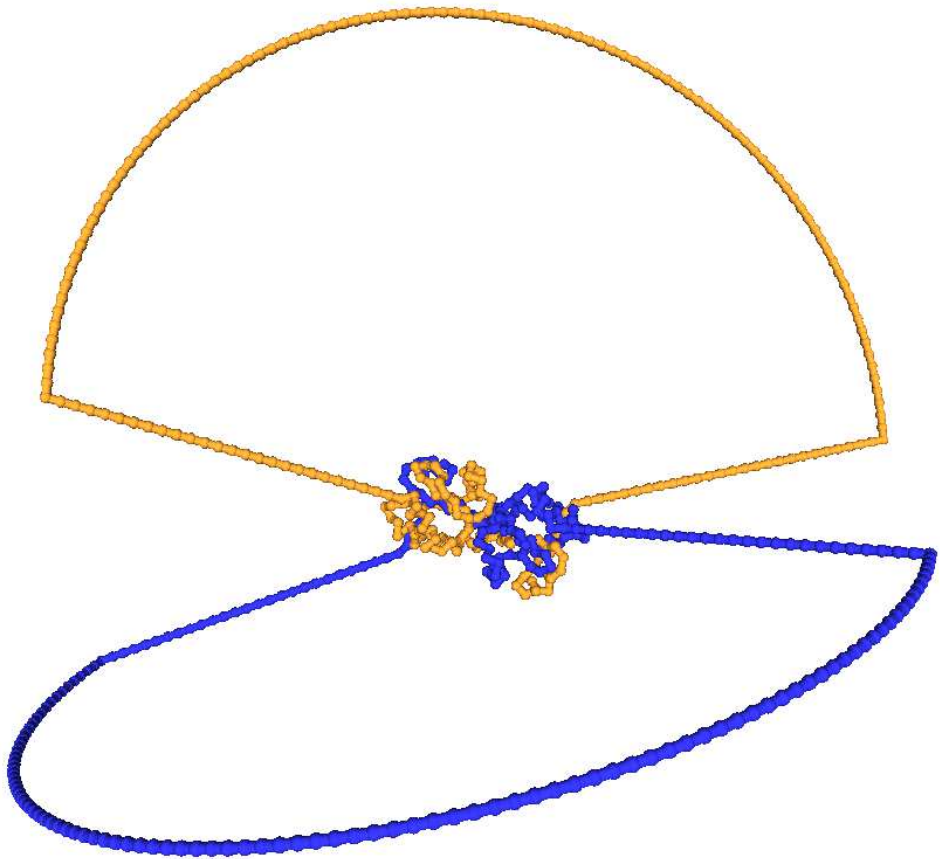} 
\caption{{\bf Example of closure in two loops}
One of the closures of the 3DIE protein. The other closures of this dimer
correspond to different orientations of the semi-circles, hinged to 
the straight, fixed segments.
\label{fig:loop}}
\end{figure}

Within the protein databank we have found $n_D =110$ non-redundant 
domain-swapped proteins and for each of them we have computed the average linking number
$G$ and the Gaussian entanglement $G'$ for open chains
(see Methods for details, and below). The results are reported  
in Table~\ref{tab:G1G}, ranked for increasing $G'$.
The table also indicates whether the dimer is human 
($33$ cases, tagged with a ``h'') or not, 
and reports the number $N$ of $C_{\alpha}$ atoms of each protein forming the dimer.

\subsection*{Protein mutual entanglement estimators: $G$ e $G'$}
 	
As a first indicator of the mutual entanglement of two proteins
belonging to a given dimer, we consider the Gauss integral $\cal G$
computed over the pair of loops obtained by closing randomly each
protein $C_\alpha$ backbones on a sphere (see Figure~\ref{fig:loop}
and Methods for details). For a given closure, $\cal G$ is an
integer~\cite{rolfsen1976} but, once averaged over several random
closures, its mean value $G$ is eventually a fractional number.

\begin{figure}[!t] 
\includegraphics[width=4in]{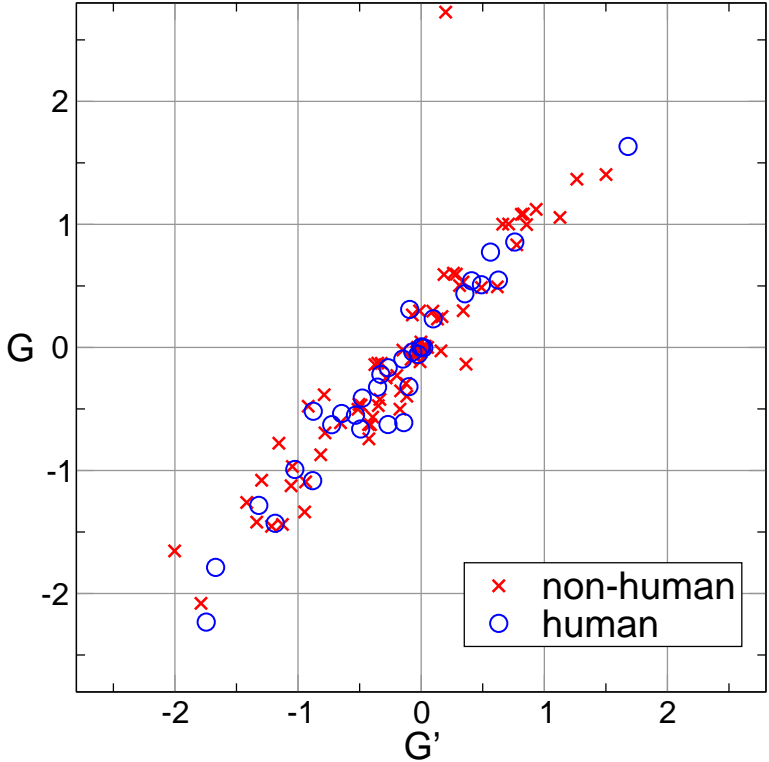} 
\caption{{\bf Correlation between the two measures of entanglement.}
Except for an outlier point, data show a good linear correlation between $G$ and $G'$.
\label{fig:GG'}}
\end{figure}

Alternatively, we can  
apply the use of the Gauss integrals to open backbones.
This measure, $G'$,  is certainly not a topological 
invariant anymore, but nevertheless it captures the interwinding between the two strands and 
does not require averages over many random closures.
By computing these two quantities on the whole set of swapped domains in our 
dataset, we can notice that there is a strong correlation between $G$ 
and $G'$ (see Figure~\ref{fig:GG'}). This result validates, at least for the domain-swapped proteins,
the use of  $G'$ as a faithful measure of the mutual entanglement. 
The reason to prefer $G'$ is twofold: First, the estimate of $G'$ does not 
require a computationally expensive averaging over different closure modes.
Second, there are cases 
in which the closure procedure does not work properly, giving rise to an
unreasonable value of $G$ (compared to  $G'$, see the point with $G>2$ and $G'\approx 0$
in Figure~\ref{fig:GG'}). 

\begin{figure}[!t] 
\includegraphics[width=4in]{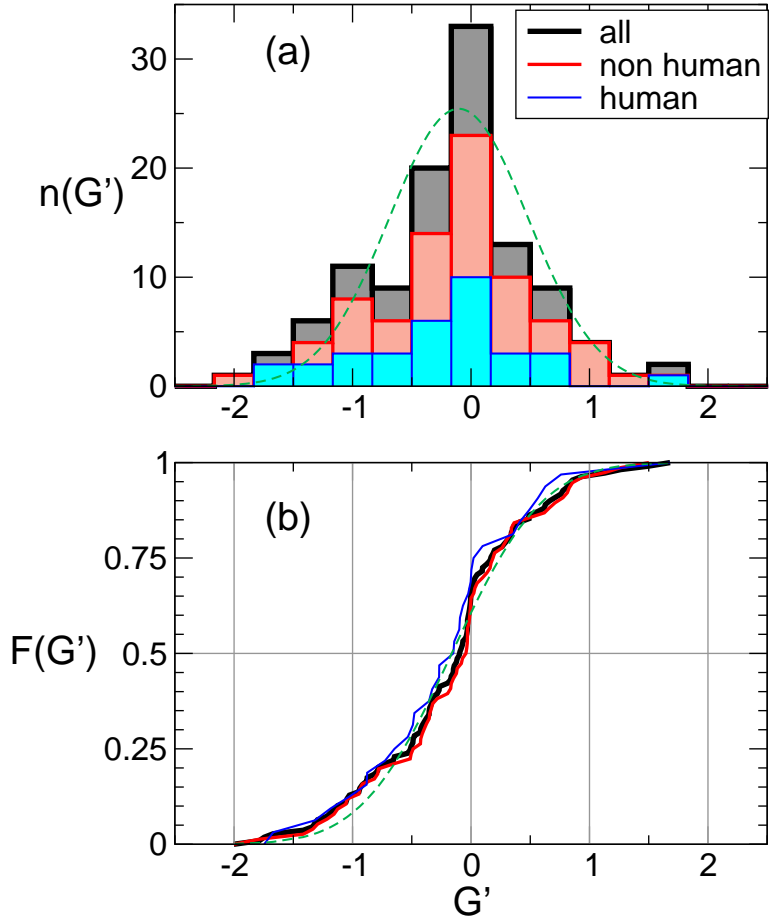} 
\caption{
{\bf Distribution of the entanglement.}
(a) Histogram of the values of $G'$
for all swapped dimers in the database, and contributions from 
the groups of human and non-human dimers.
The global histogram is fitted by a Gaussian distribution 
(dashed line) with mean  
$m\approx -0.1$ and standard deviation  $\sigma\approx 0.63$.
(b) 
Cumulative distributions of $G'$
for the same ensembles (solid lines) and an error function
fit of the the global set of data (dashed line).
\label{fig:hist} }
\end{figure}

\subsection*{Analysis of the Gaussian entanglement $G'$}

The histogram reproducing the number of
swapped dimers with a given $G'$ is plotted in Figure~\ref{fig:hist}(a):
The distribution of $G'$ is fairly well fitted by a Gaussian with 
mean $\approx -0.1$ and standard deviation $\approx 0.63$. 
The plot has 
high fraction of cases with $-1 < G' < 1$,
suggesting that most of the 3D domain-swapped dimers are not linked. 
On the other hand, there is a consistent percentage
of structures that exhibit a non trivial value of $|G'|$.
In particular, in Table~\ref{tab:G1G} 
there are $15$ structures with $G'< -1$ and $4$ dimers have $G'>1$.
Hence, more than $15\%$ of the dimers in
our databank have $|G'|$ higher than $1$.

The figures also tell us that the statistics of mutual entanglement 
displays an asymmetry towards more negative values of $G'$.
Indeed, in Table~\ref{tab:G1G} one could notice that about two thirds of the 
dimers have $G'<0$. 
For obtaining a better evaluation of the spread and average value of $G'$,
we analysed the empirical
cumulative distribution function $F(G')$, namely the fraction of
configurations that have a value at most equal to $G'$, see solid
lines in Figure~\ref{fig:hist}(b). These have been fitted by an
error-function with non-zero average. The fit yields
average $G'_0 = -0.163$ and standard deviation $\Delta G' = 0.853$
(the corresponding fit is shown as a dashed line in Figure~\ref{fig:hist}(b)).
If the fit is restricted to non-human dimers, we get
$G'_0 = -0.130$ and  $\Delta G' = 0.830$, while for the
human dimers we get   $G'_0 = -0.237$ and $\Delta G'  = 0.839$.
Again, mean values are lower than zero
both for the case of human proteins and non-human ones.  
A fraction of data $\approx 64 \%$ (non-humans) and
$\approx 70\%$ (humans) have $G'<0$. 

The asymmetry in the distribution in favor of structures with negative
$G'$ is slightly more marked for human swapped dimers.  This could be
explained by the fact that, for human proteins, the interface between
the secondary structures of the two monomers is mainly formed by swapped
$\beta$-structures: the preferential right-handed twist of the
$\beta$-sheets together with the higher frequency of anti-parallel
pairings may imply a negative value for $G'$.

\begin{figure}[!tb] 
\includegraphics[width=4in]{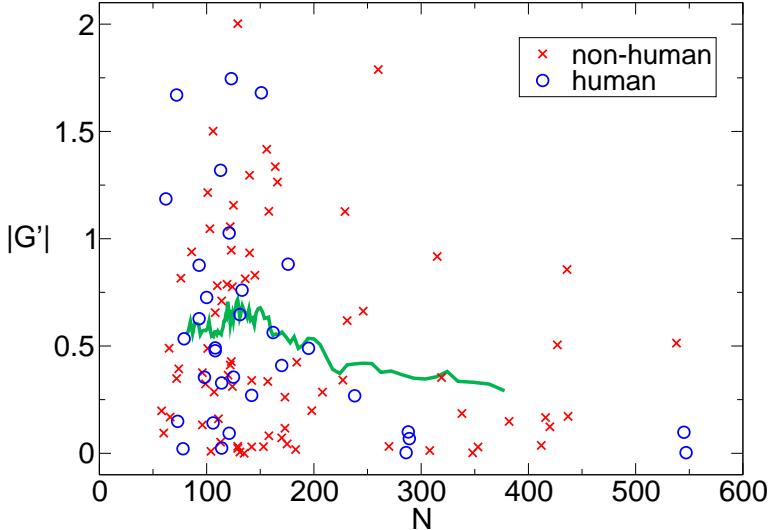} 
\caption{
{\bf Modulus of the entanglement vs.~protein length.}
Absolute Gaussian entanglement $G'$ as a function of the number of amino acids in one protein of a dimer.
The line represents a running average over $21$ points.
\label{fig:absG'} }
\end{figure}

In order to verify whether our measure of mutual entanglement is
affected by some bias that can be introduced by the different lengths
of proteins, in Figure~\ref{fig:absG'} we plot $G'$ as a function of the number of
amino acids in the proteins forming the dimers. As a matter of fact we
see that the mean value of $|G'| \approx 1/2$ is not varying significantly in the
range under investigation, which includes protein lengths ranging from about 
$50$ to $1000$ amino acids. Only fluctuations are larger at small length due to
the presence of more data.  Therefore we conclude that, in our dataset, 
the Gaussian entanglement is a parameter intrinsic to the dimers and it is not
affected by entropic effects induced by the length of the protein.

\subsection*{Dynamical entanglement}

The values of $G'$ are easy to compute from configurations and thus
represent a basic indicator of the mutual entanglement of a structure.
Through visual inspection of configurations, as those shown in
Figure~\ref{fig:ex1} and Figure~\ref{fig:ex2}, one verifies indeed that
dimers with large $|G'|$ are quite intertwined.
However, from the same figures, it appears that, in addition to a global twisting
of one protein around the other, $G'$ may be affected by some local details
of the chains, such as their 3D shape near their ends. 
These local details should be irrelevant if one thermally excites the dimer,
which should unfold with a time scale that corresponds to the Rouse dynamics
 needed to untwist one whole chain from the other \cite{doi-edwards}.

Motivated by these observations, to complement $G'$ we tackle the
problem of the entanglement from a dynamical perspective.  For some test
dimers we monitor the evolution of the value of $G'(t)$ with time, in
a Langevin simulation where only excluded volume effects play a role
(besides of course the chain connectivities). This is equivalent
to the unfolding at a sufficiently high temperature.

\begin{figure}[!tb] 
\includegraphics[width=4in]{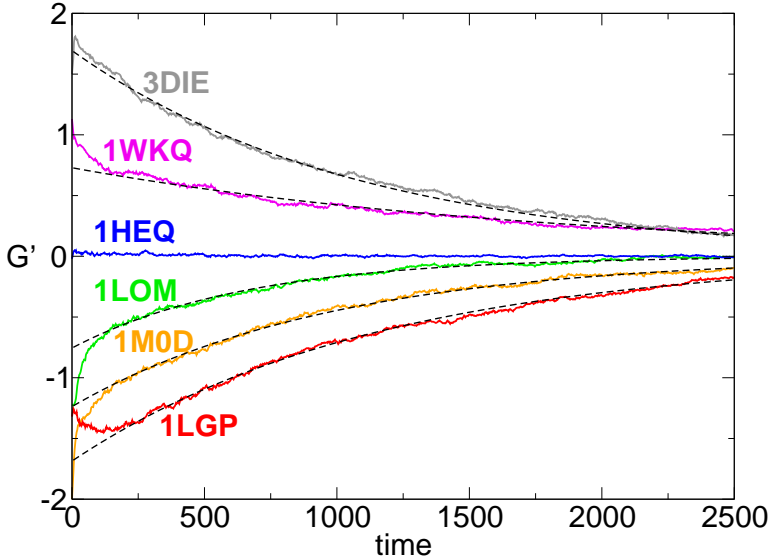} 
\caption{
{\bf Decay of the entanglement with time, during  unfolding.}
Time dependence of  $G'$ during the unfolding dynamics of a set of swapped dimers.
The dashed curves correspond to the best fit of the data 
with the function $G^*\exp(- t/\tau)$. Note that, to better catch the exponential decay, the
first 100 time steps have been neglected in the fit.
\label{fig:G'_vs_time} }
\end{figure}

 The average curve of
$G'(t)$ over many trajectories starts from the static value of the
crystallographic structure ($G'(0) = G'$) and decays to zero for long
time. It turns out that an exponential form $G^* e^{-t / \tau}$
represents well the long time decay of $G'(t)$. However, the
extrapolated value of the fit at time $t=0$, namely $G^*$, does not
necessarily match the static value $G'$.  In the studied cases, shown
in Figure~\ref{fig:G'_vs_time} and listed in Table~\ref{tab:tau}, we
find both instances of $|G^*| > |G'|$ (3DIE and 1LGP), and $|G^*| <
|G'|$ (1WKQ, 1LOM, 1M0D).  
This shows that $G^*$, a more time
consuming option than $G'$, may however be considered for complementing the
quick, static evaluation of the Gaussian entanglement.
Of course, any dynamical procedure provides a result 
that depend on the kind of dynamics used
to disentangle the structure.
For example, at room temperature 
we may expect different $G^*$'s if we perform all atom molecular
dynamics\cite{cossio2012} or Monte-Carlo methods based on local 
moves\cite{zamuner2014} (where effective interactions among
amino-acids are taken into account). 
Hence, the analysis of dimers' intertwining through the value of $G'$
may, and should, be complemented by alternative dynamical methods,
in order to get a detailed picture of the entanglement conditions.

The second parameter of the fit, the timescale $\tau$ (in simulation units), 
should increase with the chain length. 
This is expected for Rouse dynamics of polymers in general~\cite{doi-edwards}:
to shift the polymer center of mass of one radius of gyration one needs
to wait a time $\sim N^{2\nu+1}$ with Flory exponent $\nu\approx 3/5$.
It is not possible for us to asses if this scaling is respected, 
given the few cases analysed.
However, these cases correctly display an almost monotonically increasing trend
of $\tau$ with $N$ (compatibly with error bars, see Table~\ref{tab:tau}).
Note that strong logarithmic corrections to the scaling $\tau\sim N^{2\nu+1}$ are also 
expected in unwinding processes~\cite{walter2013,walter2014}.

\begin{table}[!bt]
\caption{\bf Relaxation time $\tau$ and entanglement indicators, for some swapped dimers.}
\begin{tabular}{|r|r|r|r|r|r|}
\hline {\bf dimer} & $N$    & $\tau$      & $G^*$   & $G'$    & $G$\\
\hline 1MOD  & $129$  & $1000 (70)$ & $-1.25$ & $-2.00$ & $-1.65$ \\
\hline 1LGP  & $113$  & $1140 (80)$ & $-1.70$ & $-1.33$ & $-1.28$ \\
\hline 1LOM  & $101$  & $660  (80)$ & $-0.77$ & $-1.23$ & $-1.46$ \\
\hline 1WKQ  & $155$  & $2500 (120)$& $0.72$  & $1.13$  & $1.07$  \\
\hline 3DIE  & $107$  & $1090 (90)$ & $1.68$  & $1.50$  & $1.41$  \\
\hline
\end{tabular}
\label{tab:tau}
\end{table}

\section*{Discussion}

Mathematically, two curves are linked or not, in a rigorous sense, only if
they are loops.
Hence, it is not trivial to estimate the level of intertwining between two open chains. 
Yet, the mutual entanglement is a well-visible feature in the
crystallographic structures of several domain-swapped dimers.  Being
interested in quantifying such entanglement, with
Gauss double integrals over the backbones of the two proteins  in
the dimers we provide a simple and efficient indicator, the Gaussian entanglement $G'$.
Indeed, according to our comparisons, a procedure for looping
each protein (with an artificial continuation escaping from
the core of the dimer) produces on average a linking
number $G$ that is strongly correlated with $G'$. This suggests that 
such procedure can be avoided in practice, one may just rely on the
information from the open chains, encoded in $G'$.

We report that about $15\%$ of the domain-swapped dimers have a
significant $|G'|>1$.  This is  quite intriguing, especially if
compared with corresponding  figures for knotting of single proteins, 
where about $1$\% of the PDB entries has been found to host a 
knot~\cite{jamroz2014}.

The asymmetry in the typical values of $G'$, with many more cases with
$G'<1$ than with $G'>1$, is another interesting feature emerging from
our analysis. This asymmetry could be explained by the conjecture that
the Gaussian entanglement, despite being a global feature, can be
deduced from the local twisting of closely interacting swapped
structural elements. In several cases the latter are $\beta$-strands
within the same sheet (see for example 1MOD, 1LGP in Figure 1 and
3DIE, 1WKQ in Figure 2), so that the more frequent case of a
right-handed anti-parallel $\beta$-sheet \cite{nowick2013} would indeed
imply a negative Gaussian entanglement.  With a preliminary overview, we note
that anti-parallel swapped $\beta$-sheets are indeed occurring more
frequently in the human dimers of our database than in the non-human
ones, and human dimers have indeed on average a more negative $G'$.
The dependence of the intertwining of domain-swapped protein dimers
on the local twist of swapped interacting elements is a feature
clearly worth further investigation. If confirmed, the tuning of local
interactions could then be an evolutionary mechanism used by natural
selection to control the emergence of topological entanglement in
domain-swapped dimers.
We also tried to investigate whether there is a correlation between
entanglement and pathological states. However, our analysis
of domain-swapped dimers associated
with pathologies~\cite{shameer2012} does
not show the emergence of any clear pattern.

The longer the proteins in the dimers, the slightly lower is their
mutual entanglement.  This is a surprising feature because one might anticipate that
longer chains should be more easily entangled than shorter ones. Thus, it seems
that the natural selection has acted against a form of random interpenetration
in long domain-swapped protein dimers.

Via numerical simulations of the dissociation of
some dimers, we observe that the presence of a non-trivial mutual
entanglement is a robust characteristic, which vanishes exponentially
with time during the dimer unbinding. The exponential fit furnishes
a characteristic disentanglement time $\tau$ whose values does not
depend on $G'$ and is weakly correlated with the length of the
proteins. The amplitude of the exponential decay of the Gaussian entanglement
furnishes a further estimate ($G^*$) of the intertwining in the dimer,
which complements the $G'$ of the crystallographic structure (they are
not exactly equal to each other) in assessing the amount of linking in
the dimer.

Our new approach of classifying domain-swapped protein dimers
according to their mutual entanglement will likely add novel insight
on the crucial role played by the generic topological properties of
linear polymer chains in the protein context. As already demonstrated
in the case of knotted protein folds~\cite{lim2015}, the presence of a global
topological constraint, such as the linking between two protein chains, may
strongly impact the conformational properties, the thermodynamic and kinetic
stability, the functional and evolutionary role of domain-swapped
structures.

Finally, it is interesting to speculate on the possible outcome
of a single-molecule experiment performed by pulling apart the two
protein backbones of a domain-swapped dimer with significant entanglement
(high $|G'|$).  A similar experiment was carried on very recently
for single-domain protein knots, showing that the knotting topology
of the unfolded state can be controlled by varying the pulling
direction~\cite{rief2016}.  In the linked dimer case, similarly, we expect
the choice of the pulling directions to be crucial in allowing or not
dimer unlinking and dissociation into monomers.

\section*{Methods}

\subsection*{Data bank of 3D domain-swapped dimers}

In order to derive a statistically significant ensemble of
non-redundant domain-swapped dimers, we merge two existing databanks of
domain-swapped proteins, namely {\em
  3Dswap}\cite{shameer2011,shameer2011b,3Dswap} and {\em
  ProSwap}\cite{turinsky2011,proswap}.  These databanks provide
curated information and various sequence and structural features about
the proteins involved in the domain swapping phenomenon. PDB entries
involved in 3D Domain swapping are identified from integrated
literature search and searches in PDB using advanced mining options.
We first consider only the dimers and, to avoid the presence of
related structures, we consider the UniprotKB code.  For each code, we
select only one protein, the one obtained experimentally with the
highest experimental accuracy. We then filter the remaining structures
to avoid the presence of holes in the main backbone
chain. Specifically, we discarded those proteins in which the distance
between two $C_{alpha}$, listed consecutively in the pdb file, is
bigger than $10$~\AA. As a matter of fact, such holes could affect
dramatically the reliability of the linking number and there is not
an obvious strategy to join them artificially.  At the end we obtained
a databank of $n_D=110$ proteins, whose PDB code is reported in
Table~\ref{tab:G1G}. Out of these, $33$ were human.

\subsection*{Gauss integrals}
Our procedure for estimating the amount of linking between two proteins uses Gauss integrals.
As representative of the backbones of proteins, we consider
the chains connecting the coordinates $\vec r$ of the $C_\alpha$ atoms of amino acids,
which are $N$ in each of the two proteins in the dimer.

A definition of linking number between two closed curves $\gamma_1$ and $\gamma_2$ in $3$ dimensions
is given by the Gauss double integral,
\begin{equation}
\label{Gauss-int}
{\cal G} \equiv \frac{1}{4 \pi} \oint_{\gamma_1}\oint_{\gamma_2}
\frac{\vec r^{(1)} -\vec r^{(2)}}{\left| \vec r^{(1)} - \vec r^{(2)}\right|^3} 
\cdot (d \vec r^{(1)} \times d \vec r^{(2)})
\end{equation} 
This formula yields integer numbers ${\cal G} = 0, -1, 1,-2,2,$ etc.
Such definition is adapted to compute the amount of linking between
two proteins. We need first to define a procedure that closes each open chain
via the addition of artificial residues.  This closure starts by
computing the center of mass of the dimer and by considering such
center as a repeller for the new growing arms. Let us describe the
method for protein $1$, as for protein $2$ it is exactly the same:
from each of the two free ends we continue with a path expanding
diametrically from the center of mass, with a length corresponding to 
$n = 25$ typical $C_\alpha$ - $C_\alpha$ distances $\ell\approx 3.8 $ \AA.  
At this stage the polymer is composed by $N+2 n$
residues.  Since there is some arbitrariness in the final closure
joining the two artificial new end residues, we perform a statistics
over $12$ different closures, each being a meridian along a sphere
where the poles are the artificial end residues. Semicircular closure
paths are drawn, each containing a number of artificial residues $n'$
that makes their bond distance as close as possible to $\ell$.  
An example is shown in Figure~\ref{fig:loop}.

Each closed chain becomes a collection of $N_{\rm tot} = N+2 n+n'$ points
$\{\vec r^{(1)}_1, \vec r^{(1)}_2, \ldots,\vec r^{(1)}_N, \vec r^{(1)}_{N+1}, \ldots, \vec r^{(1)}_{N_{\rm tot}}\}$ 
separated by about a fixed spacing 
$|\vec r_{i+1} - \vec r_{i}|\approx \ell$, so that the
integrals are replaced by sums over segments $d \vec R^{(1)}_i = \vec
r^{(1)}_{i+1} - \vec r^{(1)}_i$, for which we use the midpoint
approximation 
$\vec R^{(1)}_i \equiv (\vec r^{(1)}_{i+1}+\vec r^{(1)}_i)/2$.  
For a given choice $z$ of the closure for both
proteins, out of the ${\cal Z}=12 \times 12=144$ possible ones, we have
\begin{equation}
{\cal G}_{z} \equiv \frac{1}{4 \pi} \sum_{i=1}^{N_{\rm tot}}
\sum_{j=1}^{N_{\rm tot}} \frac{\vec R_i^{(1)} - \vec R_j^{(2)}}{\left|
  \vec R_i^{(1)} - \vec R_j^{(2)}\right|^3} \cdot (d \vec R_i^{(1)}
\times d \vec R_j^{(2)})
\end{equation} 
(indices run periodically, hence $N_{\rm tot}+1 \to 1$, and the notation leaves the dependence of 
${\vec R}$'s on $z$ understood), and the final estimate of linking is an average over ${\cal Z}$ closures
\begin{equation}
G = \frac 1 {\cal Z} \sum_{z=1}^{\cal Z} {\cal G}_z \,.
\end{equation}

A closure provides an integer linking number ${\cal G}_z$, though the final
average $G$ may become not integer if closures with different ${\cal G}_z$ are generated.
As an alternative, we relax the requirement to have basic integer indicators of linking and 
we perform the double Gauss discrete integral over the open chains. The Gaussian entanglement indicator
\begin{equation}
G' \equiv \frac{1}{4 \pi} \sum_{i=1}^{N-1} \sum_{j=1}^{N-1} 
\frac{\vec R_i^{(1)} - \vec R_j^{(2)}}{\left| \vec R_i^{(1)} - \vec R_j^{(2)}\right|^3} 
\cdot (d \vec R_i^{(1)} \times d \vec R_j^{(2)})
\end{equation} 
has no statistical averaging and is a straightforward alternative to $G$ in the estimate of the linking of proteins.

\subsection*{Simulations}
In our
molecular dynamics simulations, each protein forming the dimer is modeled as a self-avoiding chain 
of $N$ beads. Each bead has radius $\sigma$ and is centered in the $C_{\alpha}$ position of a residue.
Adjacent beads of each protein are tethered together into a polymer chain by 
an harmonic potential with the average $C_{\alpha}$-$C_{\alpha}$
distance along the chain equal to $1.5\sigma$.
To take into account the  excluded volume interaction between beads we consider the truncated Lennard-Jones 
potential

\begin{equation}
U_{\rm \small LJ} = \sum_{i,j>i}^N 4 \epsilon \, \left[ \left( \frac{\sigma}{d_{i,j}} \right) ^{12} - \left( \frac{\sigma}{d_{i,j}} \right) ^6 + \frac{1}{4} \right] \theta(2^{1/6} \sigma- d_{i,j} ) 
\end{equation} 
where $d_{i,j} = \left| \vec{r}_i-\vec{r}_{j} \right|$ is the distance of the bead centers $i$ and $j$, $\theta$ is the Heaviside function and $\epsilon$ is the characteristic unit of energy of the system which is set equal to the thermal energy $k_B T$.
  
The system dynamics is described within a Langevin scheme:
\begin{equation}
m \ddot{\vec{r}}_i = - \gamma \dot{\vec{r}}_i - {\nabla U_i} + \vec{\eta}_i
\label{lang}
\end{equation}
where $U_i$ is the total potential of the $i$th particle, $\gamma$ is the friction coefficient
and $\eta$ is the stochastic delta-correlated noise.
The variance of each Cartesian component of the noise, $\sigma_{\eta}^2$ satisfies the usual fluctuation dissipation relationship $\sigma_{\eta}^2 = 2\gamma k_B T$.
As customary, we set $\gamma = m/(2 \tau_{LJ})$, with $\tau_{LJ}= \sigma \sqrt{m/\epsilon}= \sigma \sqrt{m/k_B T}$ 
being the characteristic simulation time. From the Stokes friction coefficient of spherical beads of 
diameters $\sigma$ we have $\gamma = 3 \pi \eta_{sol} \sigma$, where $\eta$ is the solution viscosity.
By using the nominal water viscosity, $\eta_{sol}=1$cP and setting $T=300$K and $\sigma=2.5$nm, one has $\tau_{LJ} = {6 \pi \eta_{sol} \sigma^3/\epsilon} = 74$ns.

To study the unfolding dynamics of a given dimer we take its folded configuration, as given by the PDB, as the initial 
condition.  For each initial condition we generate 100 different molecular dynamics trajectories by 
integrating numerically~(\ref{lang}) up to $t=10^4 \tau_{LJ}$.
During the dynamics we monitor the quantities $G$ and $G'$. In Figure~\ref{fig:G'_vs_time} the curves 
are obtained by averaging over the $100$ trajectories. 
Simulations are performed with the package LAMMPS~\cite{LAMMPS}.

 \section*{Acknowledgments}
EO and AT acknowledge financial support from MIUR (Ministero Istruzione Università Ricerca) through Programmi di Ricerca Scientifica di Rilevante Interesse Nazionale,
2010HXAW77\_011 (URL prin.miur.it).
AT acknowledges financial support from the University of Padua (Università degli Studi di Padova, URL www.unipd.it) through Progetto di Ateneo CPDA121890/12.
We thank two anonymous Referees for their useful comments and one of them for
suggesting the name {\it Gaussian entanglement} for the indicator $G'$.




\end{document}